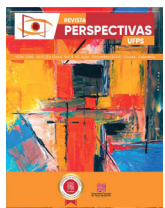
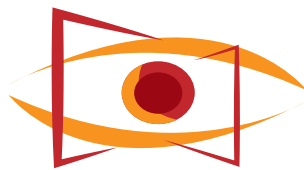

**Original Article**

# Física, Ambiente y Educación Ambiental; Percepciones desde los docentes de Ciencias Naturales en formación

Physics, Environment and Environmental Education; Perceptions from trainee Natural Science teachers

Daniel Alejandro Valderrama[1*], Marlon Damian Garzón Velasco[2], Lina Paola Alfonso Chaparro[3]

[1*]*Licenciado en Ciencias Naturales y Educación Ambiental, daniel.valderrama@uptc.edu.co, ORCID: https://orcid.org/0000-0002-3360-3890, Universidad Pedagógica y Tecnológica de Colombia, Tunja, Boyacá, Colombia.*
[2]*Estudiante de Licenciatura en Ciencias Naturales y Educación Ambiental, marlon.garzon@uptc.edu.co, ORCID hhttps://orcid.org/0009-0001-6687-0426, Universidad Pedagógica y Tecnológica de Colombia, Tunja, Boyacá, Colombia.*
[3]*Estudiante de Licenciatura en Ciencias Naturales y Educación Ambiental, lina.alfonso03@uptc.edu.co, ORCID https://orcid.org/0009-0001-2601-0667, Universidad Pedagógica y Tecnológica de Colombia, Tunja, Boyacá, Colombia.*

## RESUMEN

**Palabras clave:**

Educación Ambiental, Formación Docente, Física, interdisciplinariedad, Percepciones

La Educación Ambiental (EA) es vital para formar ciudadanos que entiendan y valoren la sustentabilidad con alternativa epistemológica y practica que mitiga las problemáticas ambientales actuales. Esta investigación surge a partir de la problematización de la relación entre la EA y las ciencias físicas, integraciones que frecuentemente se pasa por alto en los currículos y en los procesos de enseñanza de esta ciencia y de la EA. Es crucial resaltar que la física brinda marcos conceptuales y herramientas metodológicas que pueden profundizar la comprensión de fenómenos ambientales desde una óptica amplia y multidimensional. Para explorar estas relaciones, se llevó a cabo una investigación con un matiz hermenéutico interpretativo. A través de un cuestionario, se recogieron las percepciones de los futuros docentes en el área de ciencias naturales sobre el tema. Los hallazgos revelaron que una considerable cantidad de ellos reconoce y valora la correlación entre la física y la EA. Desde su punto de vista, esta unión es fundamental no solo para tener una visión integral de las dinámicas ambientales, sino también para fomentar en los estudiantes un pensamiento crítico, articulado y fundamentado frente al equilibrio ambiental. La investigación también destacó las oportunidades didácticas que se presentan al entrelazar la física con la EA. Al relacionar conceptos físicos con problemáticas ambientales reales, se puede reforzar el aprendizaje, haciéndolo significativo y perdurable en el tiempo. Esta fusión interdisciplinaria también tiene el potencial de incrementar la motivación e interés de los estudiantes, impulsando una actitud más activa y comprometida en su proceso educativo.

## ABSTRACT

Environmental Education (EE) is vital for shaping citizens who understand and value sustainability as an epistemological and practical alternative to mitigate current environmental issues. This research was prompted by the exploration of the relationship between EE and the physical sciences, connections that are often overlooked in curriculums and in the teaching processes of both this science and EE. It is essential to emphasize that physics provides conceptual frameworks and methodological tools that can enhance the understanding of environmental phenomena from a broad and multidimensional perspective. To delve into these connections, a study with a hermeneutic interpretative nuance was conducted. Through a questionnaire, the perceptions of prospective teachers in the natural sciences field regarding this topic were gathered. The findings revealed that a significant number of them recognize and value the correlation between physics and EE. From their perspective, this linkage is not only crucial for a comprehensive view of environmental dynamics but also to encourage students to develop critical, articulated, and well-founded thinking about environmental balance. The research also highlighted the didactic opportunities presented when intertwining physics with EE. By associating physical concepts with real environmental issues, learning can be reinforced, making it meaningful and enduring over time. This interdisciplinary fusion also holds the potential to increase students' motivation and interest, fostering a more active and engaged attitude in their educational journey.

**Introducción.**

El auge contemporáneo en la educación ambiental responde a una necesidad apremiante de abordar y mitigar las crecientes problemáticas ambientales que desafían nuestro planeta. Este fenómeno no es un mero capricho temporal; refleja un cambio profundo en las prioridades y percepciones de la sociedad moderna. A medida que los efectos adversos del cambio climático y la degradación ecológica se vuelven más palpables, la comunidad global se ha sumido en una revolución verde. Esta ola ecológica no solo impulsa prácticas sostenibles y políticas respetuosas con el medio ambiente, sino que también ha cultivado una tendencia firme hacia la promoción y consolidación de una conciencia ambiental robusta entre individuos de todas las edades y culturas (Valderrama & Moreno, 2023). Es evidente que, ahora más que nunca, la sociedad reconoce la imperiosa necesidad de estar informada, comprometida y proactiva en las cuestiones ambientales.

En este contexto, es importante destacar que la comprensión, la información y la conceptualización necesarias para abordar estos problemas ambientales surgen de las interacciones en diversos ámbitos, que incluyen lo físico, químico, biológico y social, Sin embargo, en estas tendencias actuales, se enfatiza en gran medida en los aspectos biológicos y químicos del ambiente(Quintero et al., 2019), mientras que ciencias como la física a menudo toman un segundo plano en la formación de la educación ambiental.

No obstante, esto se convierte en un desafío, ya que reconocer las relaciones de la física con la educación ambiental e incluso llegar a articularlas, resulta confuso y difícil, debido a la falta de investigaciones, estudios y literatura que hablen conjuntamente de estos dos campos de la ciencia. Esta falta de investigación constituye una problemática, ya que la física desempeña un papel crucial en la comprensión de los fenómenos naturales. Asimismo, en el desarrollo de tecnologías para el uso de energías renovables, los aportes de la experimentación para la comprensión de las relaciones ambientales, la toma de decisiones frente al cambio climático y aportes de la relación de la física y el ambiente a la práctica docente (Silva et al., 2019).

Es por estas razones que, para abordar este desafío, se llevó a cabo el proceso de investigación con el objetivo de identificar las percepciones que tiene los estudiantes de la Licenciatura en Ciencias Naturales y Educación Ambiental de la Universidad Pedagógica y Tecnológica de Colombia; frente a la relación entre física y educación ambiental.

Considerando que es fundamental desarrollar estrategias que correlacione de manera efectiva la educación ambiental con la enseñanza de la física. Se debe tener en cuenta que una educación ambiental sólida debe empoderar a las personas para tomar decisiones informadas y eficaces, lo que implica promover tanto la conciencia ambiental como la comprensión de los principios físicos relacionados específicamente con las problemáticas ambientales actuales. La importancia de comprender y enseñar la conexión entre la física y la educación ambiental, a fin de enriquecer la formación científica de los estudiantes y fomentar una mayor conciencia ambiental, no solo beneficia su desarrollo académico, sino que también los prepara para enfrentar los desafíos ambientales actuales y futuros con una perspectiva científica más amplia e integradora.

Este proceso de investigación permitió la exploración y categorización de las percepciones para entender la importancia de la conexión entre la educación ambiental y la física. Los docentes en formación reconocen la relevancia de los conceptos y procedimientos físicos en el estudio de las dinámicas ambientales, además promueve un pensamiento ambiental más profundo, así mismo se identifican potencialidades didácticas en la integración de estos dos saberes.

**Materiales y Metodos**

Para el desarrollo de este artículo de investigación, se adoptó el enfoque hermenéutico interpretati-vo, debido a que este permite profundizar y establecer conexiones en la comprensión de las diferentes perspectivas subjetivas de la población y muestra seleccionada. (Carr y Kemmis, 1986). La población y muestra, consistió en estudiantes de pregrado de octavo semestre que cursan el programa de Licenciatura en Ciencias Naturales y Educación Ambiental de la Universidad Pedagógica y Tecnológica de Colombia (UPTC).

Para la recolección de datos se empleó un cuestionario diseñado para identificar las diferentes percepciones de los futuros docentes de ciencias naturales, acerca de la relación entre la física y la educación ambiental. Las temáticas que se abarcaron en el cuestionario son las siguientes: tecnologías verdes, análisis y mitigación de impactos ambientales, educación práctica y experimental, fomento del pensamiento crítico, y finalmente conciencia sobre el cambio climático.

Una vez recopilados los datos, se procedió a jerarquizar y categorizar la información de acuerdo con las unidades de análisis que se agruparon en tres principales categorías: relaciones conceptuales de la física y el ambiente, relaciones procedimentales de la física y el ambiente, finalmente, física y educación ambiental (tabla 1). Posteriormente, se llevó a cabo un análisis cualitativo de las respuestas proporcionadas en el cuestionario, lo que permitió profundizar en las perspectivas de los futuros docentes de ciencias naturales, permitiendo, ofrecer así una visión más detallada de cómo los estudiantes perciben la relación entre la física y la educación ambiental.

Completada la etapa de análisis y discusión de resultados, se identificaron los hallazgos presentes en las percepciones de los docentes en formación en el campo de las ciencias naturales, acerca de la aplicabilidad de la física en la conservación y sustentabilidad del ambiente. Además, se consideró la posibilidad de aplicar este conocimiento en la práctica docente.

**Tabla I.** Unidades de análisis

| Categoría | Descripción | Preguntas |
|---|---|---|
| Relaciones conceptuales de la física y el ambiente | Esta categoría se enfoca en la intersección entre la física y las cuestiones ambientales. Específicamente, cómo los principios y conceptos físicos pueden influir, contribuir o proporcionar soluciones a problemas y desafíos ambientales. | *¿De qué manera el conocimiento sobre física ha permitido el desarrollo de tecnologías para el uso de energías renovables?* *Identifique una problemática ambiental y mencione sobre que conceptos físicos requiere claridad para resolver dicha problemática.* |
| Relaciones procedimentales de la física y el ambiente | Explora cómo las metodologías y procedimientos de la física, como la experimentación y el análisis de datos, influyen en la comprensión y abordaje de cuestiones ambientales. | *¿Cuáles son los aportes de la experimentación en física para la comprensión de las relaciones ambientales?* *La física se fundamenta en el análisis de datos para construir modelos, leyes y teorías. ¿De qué forma estas metodologías contribuyen en los estudios sobre relaciones ambientales?* *¿De qué manera la física permite tomar decisiones frente al cambio climático?* |
| Física y Educación Ambiental | Analiza la integración de conceptos y metodologías de la física en la enseñanza y aprendizaje de temas ambientales, con el objetivo de enriquecer la práctica docente y promover una educación ambiental más profunda y significativa. | *¿De qué manera se puede llevar las relaciones entre la física y el ambiente a la práctica docente?* |

## Resultados y Discusión

La enseñanza de ciencias naturales desempeña un papel crucial en la formación de ciudadanos conscientes y comprometidos con la sustentabilidad ambiental, por lo que a continuación se presentan los resultados de este estudio relacionando las categorías de estudio encontradas, e interpretando las mismas a la luz de las posibilidades didácticas y epistemológicas que podrían emerger desde estas percepciones.

*Relaciones conceptuales de la física y el ambiente*

A partir de la revisión de la literatura, se pueden encontrar diversidad de trabajos relacionados con la conceptualización en física (Llancaqueo et al., 2003; Sánchez et al., 2005), la apropiación conceptual es una de las grandes preocupaciones de la educación en general y especialmente de la educación en ciencias, ya que desde la comprensión de estos conceptos, se pueden entender los fenómenos naturales y es desde estas comprensiones que se puede gestar la innovación tecnológica y fundamentar la reflexión frente a la naturaleza de los desequilibrios ambientales.

Dichos esfuerzos en la conceptualización sí bien son importantes, pierden relevancia cuando el estudiante o ciudadano que los adquiere es incapaz de vincularlos a contextos reales en los que debe tomar decisiones (Dumrauf & Cordero, 2020; Enrique & Freire, 2019). Es por eso por lo que se hace necesario reconocer la manera en que los conceptos adquiridos por los docentes de ciencias naturales y educación ambiental han incorporado, en las asignaturas de física abordadas en su plan de estudios, a la luz de las relaciones ambientales que se podrían gestar desde estos conceptos.

En primer lugar, se indago a los estudiantes, ¿De qué manera el conocimiento sobre física ha permitido el desarrollo de tecnologías para el uso de energías renovables?, como se aprecia en la gráfica 1, el 74% de los estudiantes logra relacionar conceptos científicos de la física con desarrollos tecnológicos de las tecnologías en energías renovables, e incluso ejemplifican la forma en la que se dan esas relaciones, E5 menciona, por ejemplo: "El conocimiento acerca de las fuerzas y la capacidad de transformar la energía ha permitido la creación de paneles solares, entre otros", E12 por su parte plantea; "La física ha permitido el desarrollo de diversas tecnologías qué han tenido en cuenta el uso de energías renovables como: la energía solar, la energía eólica, la energía hidráulica. La física ofrece una opción esencial y sostenible para poder abordar los verdaderos desafíos utilizando este tipo de energías y así, reducir el impacto ambiental, generalmente dichas innovaciones son ligadas al concepto de energía y sus transformaciones"

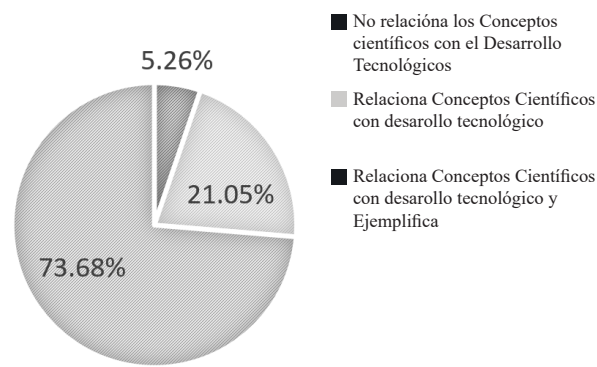

**Grafica 1.** ¿De qué manera el conocimiento sobre física ha permitido el desarrollo de tecnologías para el uso de energías renovables?

Un 21% de los docentes en formación reconocen que los avances en la comprensión y gestión de las energías renovables poseen un fundamento físico, sin embargo, no es evidente su ejemplificación o profundidad en la respuesta, paralelo a un 6% de docentes en formación que no encuentran dicha relación, esto sugiere la necesidad de pensar en procesos de formación más interdisciplinares y aplicados en la enseñanza de la física.

Complementario a lo anterior se les solicitó a los docentes en formación que reconocieran una problemática ambiental, y la relacionaran con conceptos físicos sobre los que se debía tener claridad para darle solución, con excepción de un 2% todos los docentes en formación lograron llegar a dicha asociación, logrando en el grupo, articular las respuestas en torno a las problemáticas de contaminación auditiva, atmosférica y del agua, así como al calentamiento global, la pérdida de biodiversidad y problemas ambientales; es significativo ver como los docentes asocian a la comprensión y mitigación de esas problemáticas conceptos asociados a ramas de la física como mecánica, termodinámica y los fenómenos ondulatorios, dichas relaciones se ilustran en la tabla II.

**Tabla II.** Relación de conceptos físicos y problemáticos ambientales.

| Problemática | Conceptos |
|---|---|
| Contaminación Auditiva | Acústica |
| Calentamiento Global | Termodinámica, Comportamiento de la Luz, Energía, Gases, Radiación, Temperatura. |
| Pérdida de Biodiversidad | Modelado de datos |
| Contaminación atmosférica | Mecánica de fluidos, Transformación de la energía, Fuerza |
| Contaminación del Agua | Física de Materiales, Nanopartículas |
| Desastres Naturales | Física de Materiales, Nanopartículas |

### *Relaciones procedimentales de la física y el ambiente*

La física como ciencia natural se fundamenta en el método científico y epistemológicamente se ha desarrollado a partir de estructuras conceptuales fundamentadas en la observación de los fenómenos naturales, la formulación de hipótesis, la toma de datos experimentales y el análisis de los mismos a partir de los modelismos matemáticos, una estructura que le confiere cierta objetividad y a la vez cierto dinamismo proporcional al desarrollo tecnológico y el mejoramiento instrumental derivado del mismo (Haza et al., 2023), si bien las relaciones ambientales integran aspectos muy aleatorios como la cultura o la sociedad, ante posturas sistémicas de análisis los modelos epistémicos y metodológicos de la física podría aplicarse al análisis de las principales dinámicas ecosistémicas, por ejemplo en la distribución de la energía de dichos sistemas (Márquez Delgado et al., 2021; Norman Gómez et al., 2020).

Desde la perspectiva anterior se les pide a los docentes en formación que teniendo en cuenta el fundamento procedimental de la física en cuanto al análisis de datos, para construir modelos, leyes y teorías, respondieran a la pregunta ¿De qué forma estas metodologías contribuyen en los estudios sobre relaciones ambientales?

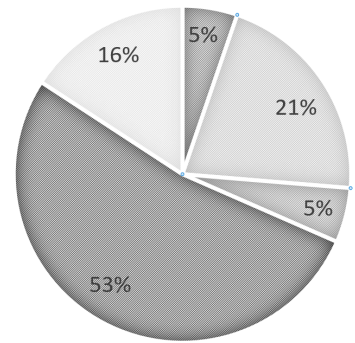

- No se reconocen relaciones posibles.
- Reconoce que lo métodos de la física contribuyen pero no explican de qué manera o ejemplifican al respecto.
- Respuesta incompleta o incomprensible
- Se reconoce y ejemplifica de manera argumentada la fisica y los estudios ambientales
- Se reconocen en forma básica argumentos sobre la contribución de los metodos de la física en los estudios ambientales

**Grafica 2**. : La física se fundamenta en el análisis de datos para construir modelos, leyes y teorías. ¿De qué forma estas metodologías contribuyen en los estudios sobre relaciones ambientales?

Como se aprecia en la gráfica 2, el 53% de los estudiantes reconoce y ejemplifica de manera argumentada la relación entre los métodos de la física y los estudios ambientales, algunas relaciones las establecen desde la necesidad de medir los fenómenos para tomar decisiones frente a los mismos, como lo plantea E8 "Mediante ecuaciones se puede tener calcular factores como el brillo solar, precipitación y otras variables para calcular el clima en diferentes estaciones y del mismo modo comparar con cambios que suceden según diversas circunstancias", otros afirman que dichos modelos pueden ser predictivos, por lo que permiten analizar variables ambientales, que definirán las perspectivas futuras de los ecosistemas, E12: "La física nos permite no solamente identificar estos cambios climáticos sino que nos lleva a utilizar la simulación de esta misma y el comportamiento en la atmósfera en diferentes puntos geográficos. Este tipo de modelos nos pueden ayudar a proyectar escenarios o situaciones futuras y de esta manera evaluar el impacto de las actividades humanas en el planeta tierra" (Sic).

Adicional a lo anterior, algunos reconocen las contribuciones de los métodos físicos en la comprensión de las relaciones ambientales, un 16%, si bien no argumentan o generan explicaciones adicionales, plantean que pueden reconocer dichas relaciones E7: "el uso de energías limpias , como tener beneficios con la energía solar por ejemplo" considerando que los métodos de la física sirven para comprender el uso de las energías limpias y obtener beneficios de las mismas, lo que en términos generales es cierto, en cuanto a que el estudio de los materiales, las relaciones luz y materia, se fundamentan en tomas de datos experimentales que se hacen faticos en el desarrollo de dispositivos como las celdas solares.

Frente a los proceso de experimentación, su utilidad es multidimensionalidad, no solo en la construcción de leyes, modelos y teorías a partir del diseño experimental y la comprobación o descarte de una hipótesis, sino también en aspectos como la innovación de dispositivos y procesos, que hacen más eficiente su desempeño, otra dimensionalidad de la experimentación tiene que ver con sus potencialidades didácticas pues se ha reconocido ampliamente que permite el desarrollo del pensamiento crítico y facilita los procesos de enseñanza y aprendizaje de las ciencias naturales (Castrillón-Yepes et al., 2020; Roberto 2021).

Con base en lo anterior, se preguntó a los docentes en formación acerca de los aportes de la investigación en física para la comprensión de las relaciones ambientales. Este es uno de los aspectos llamativos de este estudio, ya que la diferencia entre quienes ven la utilidad de la experimentación y quienes definitivamente no logran apreciar sus aportes es de apenas un 16%, como lo muestra la gráfica 3. E14 reconoce que uno de los aportes es el diseño de sistemas: "Generar diversos métodos como el método LIDAR para medir la capa de ozono y contribuir al conocimiento sobre cuánto estamos contaminando", pues estos mismos derivan de diseños experimentales particulares. E20, por su parte, menciona: "La experimentación en física proporciona datos empíricos y evidencia científica clave para comprender las relaciones ambientales, validar teorías, medir variables ambientales, identificar retroalimentaciones y desarrollar tecnologías sostenibles.

Una de las explicaciones sobre el 42% de docentes en formación que no reconocen los aportes de la experimentación, se hizo evidente en el diálogo con el grupo, donde se encontró la percepción de experimentación como una acción limitada a un laboratorio con unos recursos particulares, así como, la visión de que la experimentación siempre es de confirmación y no de descarte de hipótesis, esto sugiere la necesidad de transformar la enseñanza de la experimentación en física, hacia posturas más activas, en las que los estudiantes no se limiten a seguir los pasos de una guía sino que diseñen experimentos fundamentados en casos reales de aplicación o de confrontación de saberes.

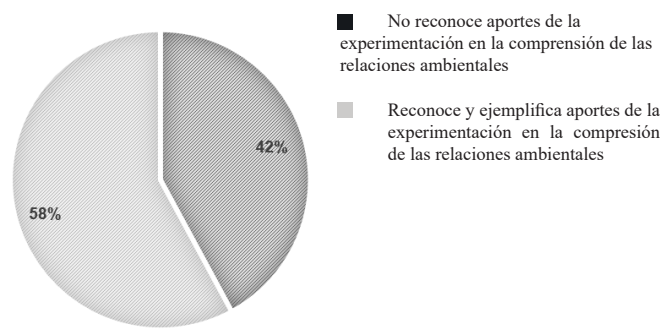

**Grafica 3.** ¿Cuáles son los aportes de la experimentación en física para la comprensión de las relaciones ambientales?

Una de las problemáticas ambientales de orden mundial, que más importancia a tomado en los discursos ambientales, tiene que ver con el cambio climático, alteración en los patrones y condiciones climáticas de la Tierra que implica variaciones significativas en las temperaturas promedio, los patrones de precipitación, los niveles del mar y otros aspectos del clima global. Este fenómeno es impulsado principalmente por la actividad humana, como la emisión de gases de efecto invernadero (como el dióxido de carbono) provenientes de la quema de combustibles fósiles, la deforestación y otras actividades industriales (Díaz, 2012)

Dichas dinámicas conllevan una gran cantidad de conceptos y métodos de la física para su interpretación y comprensión, en términos generales, losfenómenos relacionados con la luz, la reflexión, refracción, la interacción entre la luz y las moléculas de metano, dióxido y monóxido de carbono, la termodinámica y transformación de la energía entre otras, de esta manera la física ha sido una de las principales ciencias encargadas del monitoreo y la toma de decisiones frente a este fenómeno (Masson-Delmotte et al., 2021).

Desde esta perspectiva se preguntó a los docentes en formación, ¿De qué manera la física permite tomar decisiones frente al cambio climático?, ante lo cual, el 84% Reconoció la utilidad de la física frente a la toma de decisiones ambientalmente responsables en el contexto del cambio climático, E3 propone que "Si entendemos el fenómeno propiciaremos cambios en nuestra conducta, por que habremos generado procesos de aprendizaje, que en este caso deben partir desde la física" Postura que ha sido ampliamente discutida y fundamentada, en el sentido de que la comprensión y enseñanza de la física debe tener como propósito principal el desarrollo del pensamiento crítico, que implica la toma de decisiones fundamentadas frente a problemáticas como esta, (Chen-Quesada et al., 2019; Uskola et al., 2021) e incluso se ha recomendado incluir este tipo de cuestiones socio científicas en la enseñanza de la física, por la posibilidad que ofrecen de contextualizar los conceptos y privilegiar el aprendizaje activo (Torres Merchán et al., 2023). Otras respuestas son de tipo más conceptual como E15, quien plantea que "Desde la descripción de la "luz" como generadora de calor dentro de la superficie de la tierra; el comportamiento de esta tanto como onda como de partícula contribuyen a los cambios fisicoquímicos de elementos presentes en el planeta, así que partiendo de esto se ejecutan análisis como indicadores de alerta para posteriormente tomar una decisión de resolución al problema climático".

El 16% aún presentan dificultades para visualizar esas relaciones, lo que reincide en la necesidad de procesos de enseñanza y aprendizaje de la física desde perspectivas más aplicadas y bajo patrones de interdisciplinariedad con otras áreas de importancia para el desarrollo humano y natural, como la educación ambiental.

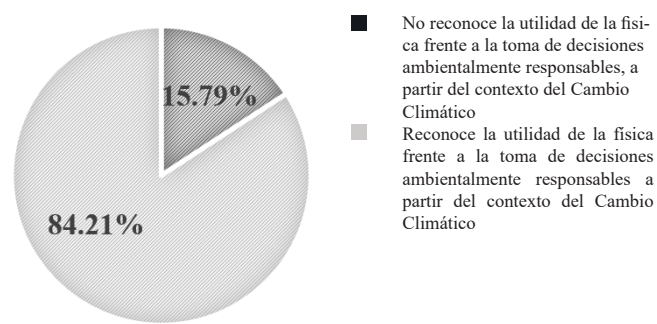

**Grafica 4.** ¿De qué manera la física permite tomar decisiones frente al cambio climático?

### *Física y Educación Ambiental*

A partir de los anteriores apartados se hizo evidente la importancia de la física en términos de comprensiones conceptuales sobre el ambiente, también en la toma de decisiones responsables y fundamentadas que propendan por el equilibrio ecosistémico y la sustentabilidad ambiental, dicha importancia se acopla perfectamente a los objetivos de la educación ambiental, que busca aumentar la conciencia y el conocimiento sobre los problemas ambientales, fomentar actitudes y comportamientos responsables hacia el ambiente, y empoderar a las personas para tomar decisiones informadas y sostenibles en relación con la naturaleza y los recursos naturales (Avendaño & Febres Cordero-Briceño, 2019).

Esta forma de educación busca proporcionar a las personas las herramientas y el entendimiento necesarios para comprender la interconexión entre los seres humanos y su entorno, así como los impactos de sus acciones en el ambiente. La educación ambiental puede abordar una amplia gama de temas, como la conservación de la biodiversidad, la gestión de recursos naturales, la contaminación, el cambio climático, la sostenibilidad, la energía renovable, dichos temas se relacionan en modo directo e indirecto con la física, por lo que la educación ambiental podría ser un buen escenario de contextualización de los conceptos físicos y su aplicación efectiva en la realidad.

Desde esta perspectiva, se les preguntó a los docentes en formación, ¿De qué manera se pueden llevar las relaciones entre la física y el ambiente a la práctica docente? Derivándose de tal pregunta la nube de palabras de la figura 1. A partir de las cuales se puede leer un discurso pedagógico constructivista, que vincula el dinamismo de las problemáticas ambientales con la contextualización de conceptos físicos, de estas palabras claves derivaron las siguientes categorías emergentes:

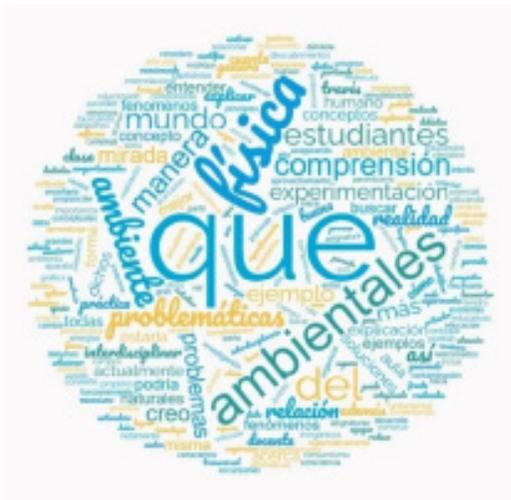

**Figura 1.** ¿De qué manera se puede llevar las relaciones entre la física y el ambiente a la práctica docente? Fuente: elaboración propia.

1. Enseñanza de la física desde la interdisciplinariedad: La interdisciplinariedad implica combinar conocimientos y enfoques de distintas disciplinas para abordar problemas complejos. Mediante la colaboración entre áreas de conocimiento, se busca obtener una comprensión más completa y soluciones innovadoras que no serían posibles desde una sola perspectiva disciplinaria (Bell Rodríguez et al., 2022).

Aunque presenta desafíos, la interdisciplinariedad puede generar avances significativos en la resolución de problemas, E20 propone "Integrando ejemplos ambientales en clases, realizando experimentos relacionados, fomentando proyectos de investigación y utilizando tecnología y excursiones educativas." Por su parte E1 plantea que "De manera que al articular estos dos campos de pueden lograr grandes descubrimientos, además de que se pueden estudiar todos los fenómenos que suceden en la naturaleza a partir de estudios realizados desde la física. Lo cual, es de gran importancia ya que se puede aplicar en el aula y así, enseñarles a los estudiantes desde una mirada interdisciplinar y acercarlos un poco más a la realidad." Esta tendencia a la interdisciplinar plantea además un reto que tiene que ver con la siguiente categoría emergente.

2. Contextualización de los conceptos físicos Una de las dificultades en la enseñanza y aprendizaje de la física que ha trascendido en el tiempo, tiene que ver con la dificultad de docentes y estudiantes para la contextualización de los conceptos, mostrándose en la mayoría de las veces de forma abstracta al punto que algunos estudiantes no le ven la aplicación en su vida cotidiana (Elizondo Treviño, 2013; Massoni & Moreira, 2010), como alternativa a esa descontextualización, los docentes en formación plantean la posibilidad de contextualizar conceptos científicos de la física en su práctica docente, e incluyen perspectivas como excursiones, experimentación, mediciones y laboratorios que permitirían reconocer problemáticas ambientales y plantear posibles solución a la luz del conocimiento científico.

En el E4, por ejemplo, plantea "Teniendo en cuenta los principios de la física se podría construir un espacio de reflexión acerca de las problemáticas ambientales, pero también se podrían construir nuevas ideas frente al aprovechamiento de los desechos ya sean orgánicos e inorgánicos partiendo del concepto de átomo, molécula y energía." En la misma perspectiva se encuentra E19, quien propone "Una excelente forma sería la experimentación creo que si se implementa se estaría asegurando una buena comprensión de los fenómenos en los educandos, además con la experimentación la atención de los estudiantes estaría enfocada netamente en el proceso y garantizaría una mejor apropiación de saberes conceptuales y prácticos, es el empirismo a lo que debemos recurrir" y E7 "Dándole un sentido a lo que enseña relacionando los conceptos con la práctica ya que como sabemos todo está relacionado con la física". Estas percepciones dejan como necesidad didáctica la articulación de la educación ambiental, en clases de física e incluso el desarrollo epistemológico de una física de carácter ambiental, cuya epistemología estaría centrada en la aplicación de los conceptos y procedimientos de la física en los estudios ambientales.

3. Pensamiento Complejo: Una perspectiva aún más ambiciosa surge desde el pensamiento complejo, un enfoque cognitivo que busca entender la realidad reconociendo la interconexión, la multidimensionalidad y la naturaleza cambiante de los sistemas y fenómenos. más allá de las visiones simplistas y lineales, abrazando la idea de que los problemas y situaciones son influenciados por una variedad de factores interrelacionados (Juárez & Comboni, 2012). Este enfoque promueve la integración de múltiples perspectivas, la consideración de contextos y la apertura a la incertidumbre. El pensamiento complejo es valioso en la resolución de problemas en ámbitos diversos, ya que ayuda a abordar situaciones intrincadas y dinámicas de manera más completa y realista, en términos didácticos el pensamiento complejo debe ser objetivo y método, pues la complejidad de los fenómenos naturales y socioculturales pueden ser asumidos únicamente mediante la integración de los saberes.

Los docentes en formación ven en la relación de la física y el ambiente, una posibilidad para desarrollar ese pensamiento complejo, pues buscan que las problemáticas ambientales sean estudiadas y mitigadas desde la complejidad, E3 menciona "Creo que en la práctica docente de ciencias naturales y educación ambiental específicamente, se hace necesario interrelacionar las distintas áreas para que los estudiantes comprendan de manera más compleja su realidad y puedan construir un criterio científico y reflexivo acerca de ello. Por lo que, por ejemplo, al explicar el aparato locomotor también se puede vincular el concepto de centro de masa, y simultáneamente los efectos que tiene el ambiente sobre el desarrollo del cuerpo humano". Visibilizando esas múltiples relaciones se asumirán enfoques de educación ambiental desde la interacción de sistemas, cambiando de esta forma visiones antropocentristas, o biocentristas del ambiente, hacia posturas de relaciones y equilibrios.

4. Pensamiento Crítico: Las relaciones entre la física y el ambiente tienen el potencial de impulsar el desarrollo del pensamiento crítico al ofrecer enfoques analíticos y conceptuales para comprender sistemas complejos y problemas ambientales. La aplicación de principios físicos a la naturaleza promueve la interdisciplinariedad, el análisis de datos, la modelización y la resolución de problemas, permitiendo una evaluación más profunda de los efectos humanos en el entorno. Esta conexión también fomenta la conciencia de sustentabilidad ambiental y la toma de decisiones informadas, sobre la cual se ha hablado anteriormente, lo que en conjunto nutre la capacidad crítica para abordar desafíos ambientales y complejos.

**Conclusiones**

Con base a los resultados obtenidos en la encuesta realizada a los estudiantes, se puede concluir que existe un alto porcentaje (74%) de estudiantes que logran relacionar conceptos científicos de la física con los desarrollos tecnológicos en energías renovables. Esto demuestra que el conocimiento sobre física ha sido fundamental en el desarrollo de tecnologías en este campo.

Además, es importante destacar que estos estudiantes ejemplifican de manera concreta cómo se han dado estas relaciones entre la física y las tecnologías en energías renovables. Por ejemplo, mencionan la creación de paneles solares y otras tecnologías que han tenido en cuenta el uso de energías renovables. Esto demuestra que la física ofrece opciones esenciales y sostenibles para abordar los desafíos relacionados con el uso de energías renovables y reducir el impacto ambiental.

Por otro lado, se observa que un porcentaje bajo de docentes en formación (21%) reconocen el fundamento físico de los avances en la comprensión y gestión de las energías renovables. Sin embargo, es importante destacar que estos docentes en formación no logran ejemplificar o profundizar en sus respuestas. De igual forma, un 6% de los docentes en formación no encuentran una relación entre la física y las tecnologías en energías renovables. Esto sugiere la necesidad de implementar procesos de formación más interdisciplinares y aplicados en la enseñanza de la física, para que los docentes puedan comprender y transmitir de manera más efectiva estas relaciones.

En cuanto a la asociación de una problemática ambiental con conceptos físicos, se destaca que la gran mayoría de los docentes en formación (98%) logran realizar esta asociación. Esto muestra que tienen claridad sobre los conceptos físicos que son relevantes para abordar y resolver problemáticas ambientales. Estos conceptos están relacionados con ramas de la física como la mecánica, la termodinámica y los fenómenos ondulatorios.

En relación con la experimentación, su utilidad es multidimensional. No solo contribuye a la construcción de leyes, modelos y teorías a partir de diseños experimentales y la comprobación de hipótesis, sino también impulsa la innovación de dispositivos y procesos para aumentar su eficiencia. Además, se reconoce que la experimentación tiene potencial didáctico, ya que facilita el desarrollo del pensamiento crítico y favorece los procesos de enseñanza y aprendizaje de las ciencias naturales.

El hecho de que el 84% de los encuestados haya reconocido la utilidad de la física como herramienta para tomar decisiones ambientalmente responsables frente al cambio climático refuerza la postura en la que se argumenta que la comprensión y enseñanza de la física deben tener como objetivo principal el desarrollo del pensamiento crítico y la toma de decisiones fundamentadas frente a problemáticas como el cambio climático. Esto implica que la enseñanza de la física debe abordar no solo los conceptos teóricos, sino también su aplicación en situaciones reales, como es el caso del cambio climático.

La interdisciplinariedad, la contextualización de los conceptos físicos y el pensamiento crítico son herramientas clave para fortalecer procesos educativos que faciliten a los docentes en formación una mayor comprensión del rol de la física en la protección y preservación del medio ambiente. De igual forma, al integrar la física en la educación ambiental, se pueden abordar problemas de manera más completa y realista, fomentando actitudes y comportamientos responsables hacia el ambiente.

## Referencias